\pdfoutput=1
\documentclass[%
 reprint,
 amsmath,
 amssymb,
 prl
]{revtex4-1}

\usepackage[dvipdfmx,dvipdfmx]{graphicx}
\usepackage{graphicx}
\usepackage{color}
\usepackage{sistyle}

\usepackage{subfigure}

\usepackage{epsfig}
\usepackage{float}
\usepackage{epstopdf}

\begin{document}

\title{Coupling colloidal quantum dots to gap waveguides
}


\author{Niels M. Israelsen$^{1,2}$ Ying-Wei Lu$^{1}$ Alexander Huck$^{1}$ and Ulrik L. Andersen$^{1}$}
\affiliation{$^{1}$ Center for Macroscopic Quantum States (bigQ), Department of Physics, Technical University of Denmark, Building 307, Fysikvej, 2800 Kgs. Lyngby, Denmark}
\affiliation{
 $^{2}$ DTU Fotonik, Department of Photonics Engineering, Technical University of Denmark, Ørsteds Plads Building 343, 2800 Kongens Lyngby, Denmark}
\email{nikr@fotonik.dtu.dk}

\begin{abstract}
The coupling between single photon emitters and integrated photonic circuits is an emerging topic relevant for quantum information science and other nanophotonic applications.   
We investigate the coupling between a hybrid system of colloidal quantum dots and propagating gap modes of a silicon nitride waveguide system. We furthermore explore the density of optical states of the system by using a scanning probe technique and find that the quantum dots couple significantly to the photonic circuit. Our results indicate that a scalable slot-waveguide might serve as a promising platform in future developments of integrated quantum circuitry.   
\end{abstract}

\maketitle

The guidance, interference and detection of light from single photon emitters on a scalable platform is of utmost importance for the development of different nanophotonic and quantum technological applications such as quantum repeaters, quantum memories and quantum computation. One of the main challenges to realize such a scalable platform is to efficiently capture the emission from single emitters onto a photonic integrated circuit in which the light guidance and interference takes place with low loss. 

Most high-efficiency demonstrations have been realized in solid state systems in which the quantum emitter is directly embedded into the solid state host, e.g. defects in diamond or gallium arsenide host materials formed as cavities or waveguides~\cite{kim2013,Hausmann2012,Evans2018}.
However, the fabrication as well as the deterministic positioning of single emitters in these systems is highly nontrivial, and it is therefore interesting to consider alternative approaches based on heterogeneous (or hybrid) systems where the emitter and photonic integrated circuit constitute different material platforms~\cite{Benson2011}.

One example is to use a metallic circuit where the single photon emitter is placed in the vicinity of a highly localized plasmonic field of a metallic structure, thereby largely changing the emission dynamics of the emitter and in some cases direct it into a propagating mode of the plasmonic system~\cite{Akimov2007, Huck2011, Kumar2013}. While significant advances have been demonstrated, e.g. a Purcell enhancement of approximately 1000~\cite{Akselrod2014}, propagation of single plasmons in small circuits~\cite{Kumar2014} and single emitter excitation of modes at the interface between metals and dielectric structures~\cite{Kumar2019}, the efficient coupling to a low-loss plasmonic system has yet to be realized. 

To minimize the losses of the hybrid system, it might be more promising to explore the coupling of emitters to dielectric waveguides. Such a coupling and subsequent photonic propagation and linear optical manipulation has been demonstrated in multiple experiments~\cite{Turschmann2017, Khasminskaya2016, Davanco2017, Elshaari2017, kim2017, Osada2019, Katsumi2019, Kim2019}.   
In this article we explore an approach to emitter-waveguide interaction based on the coupling of colloidal quantum dots to the mode formed in the gap between two Silicon Nitride (SiN) waveguides. 

 
 \begin{figure}[t]
\centering 
{
\includegraphics[width=0.9\linewidth]
{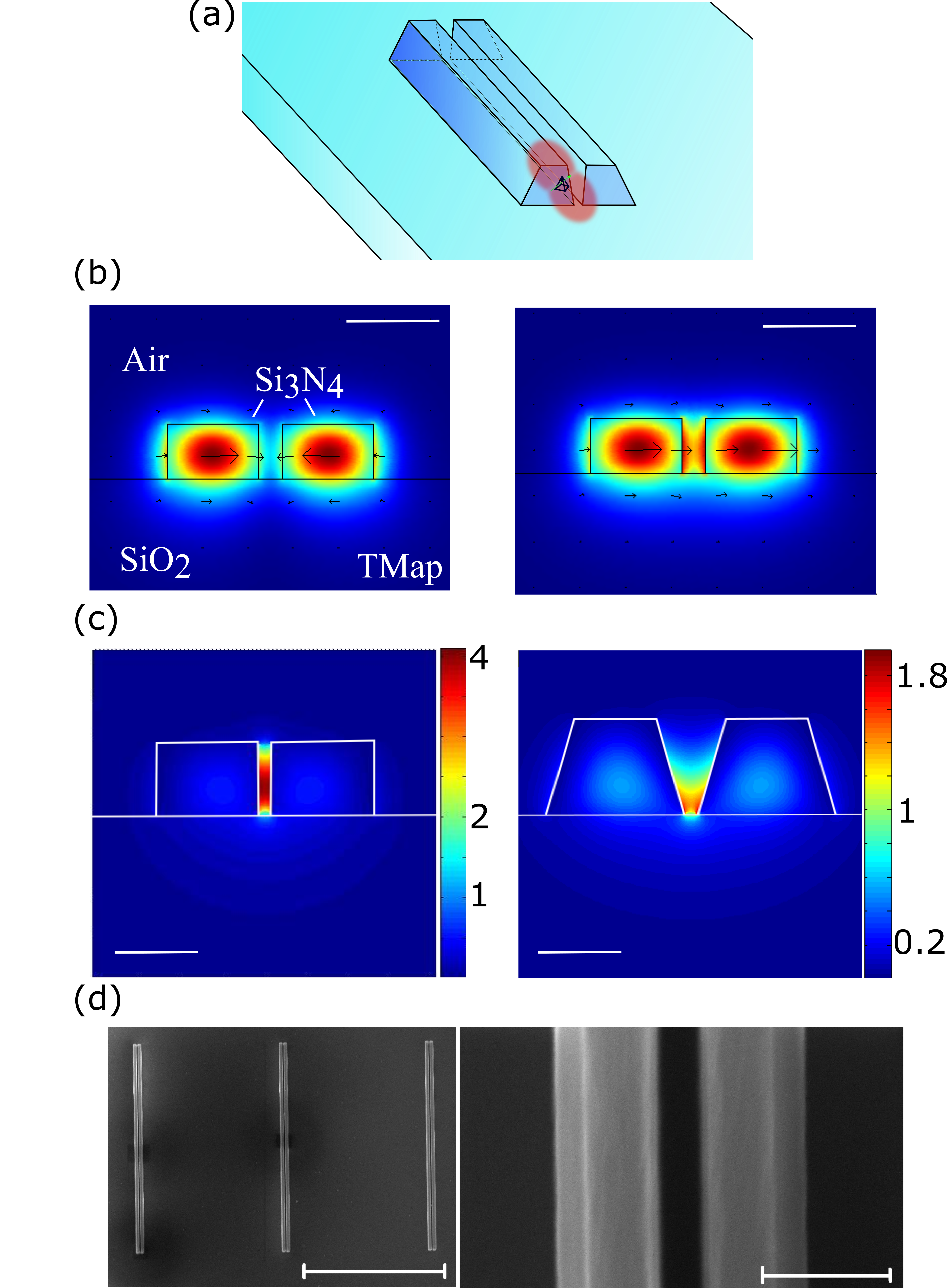}
}

\caption{Coupling emitters to a gap mode in a Silicon Nitride (SiN) waveguide system. (a) Schematic drawing of the heterogeneous system comprising a single emitter inside the gap of a waveguide. (b) Electric field, $|E|$, distribution of the two transverse magnetic waveguide modes of SiN. It is only the TM$_p$ mode (right hand side) that fills the gap. (c) Simulation result for the change in the decay rate of the emitter in the TM$_p$ mode, $\Gamma_{TM_p}/\Gamma_0$, for perfectly parallel waveguides (left) and the slanted waveguide (right).  (d) Scanning electron microscopy image of the waveguide in full size (left) and a zoom (right) where the slanted walls become visible. The scale bars are: (b) $400\,nm$, (c) $200\,nm$, (d) $10\,\mu m$ left hand side and $300\,nm$ right hand side.
\label{Fig1}
}
\end{figure}
   
The integrated platform employed in this work is schematically illustrated in Fig~\ref{Fig1}(a). It comprises an optical emitter placed in the gap of two optical dielectric waveguides where the choice of material was SiN due to its low propagation losses and its CMOS compatibility. The waveguide structure supports two transverse-magnetic (TM) photonic modes with parallel (TM$_p$) and anti-parallel (TM$_{ap}$) electric field components, as illustrated in Fig.~ \ref{Fig1}b. Only the parallel mode TM$_p$ is significantly located in the gap between the waveguides and due to this localization, we expect a strong interaction between the waveguide structure and an embedded emitter to occur in the gap. The interaction strength can be quantified by the decay rate enhancement $\Gamma_{TM_p}/\Gamma_0$ and the mode coupling efficiency $\beta = \Gamma_{TM_p}/(\Gamma_0 + \Gamma_{TM_p})$, where $\Gamma_{TM_p}$ and $\Gamma_0$ are the decay rates of the emitter when placed in the gap and in vacuum, respectively. For waveguide dimensions of $250\,nm\times 175\,nm$ (width x height) and a gap size of $30\,nm$, we obtain an enhancement of the decay rate by almost a factor of four which means that around 80\% of the emitters fluorescence will be captured by the propagating gap mode (see left plot of Fig.~\ref{Fig1}(c)~\cite{Robinson2005}. The actual fabricated waveguide geometry is however slanted as shown in the right plot of Fig.~ \ref{Fig1}c. In such a system, the mode localization in the gap decreases and hence the expected coupling rate is lower. For the dimensions shown in the figure, we can at most expect an enhancement in the decay rate of a factor about two and $\beta \approx 60\%$.

\begin{figure}[t]
\centering 
{
\includegraphics[width=1.0\linewidth]
{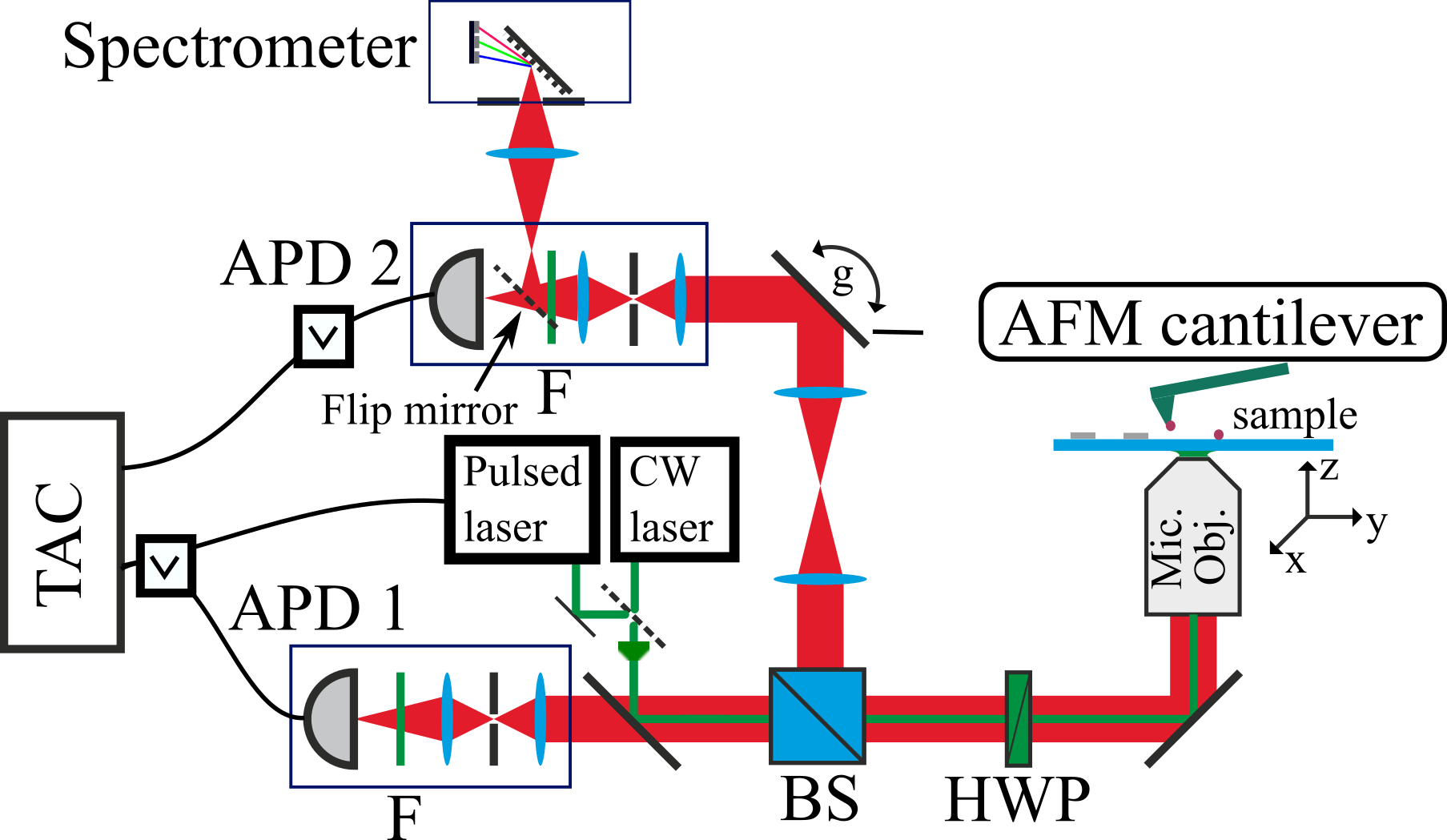}
}
\caption{Schematic of the experimental setup. See text for description. AFM: Atomic force microscope. TAC: time to amplitude converter. APD1 and APD2: avalanche photo diode detection channel 1 and 2, respectively. g:galvanometric mirror. BS:50-50 beam splitter. HWP:half waveplate. F: Notch-Filter with low transmission at $532\pm 15\,nm$.
\label{Fig2}
}
\end{figure}

\begin{figure}[t]
\centering 
{
\includegraphics[width=1\linewidth]
{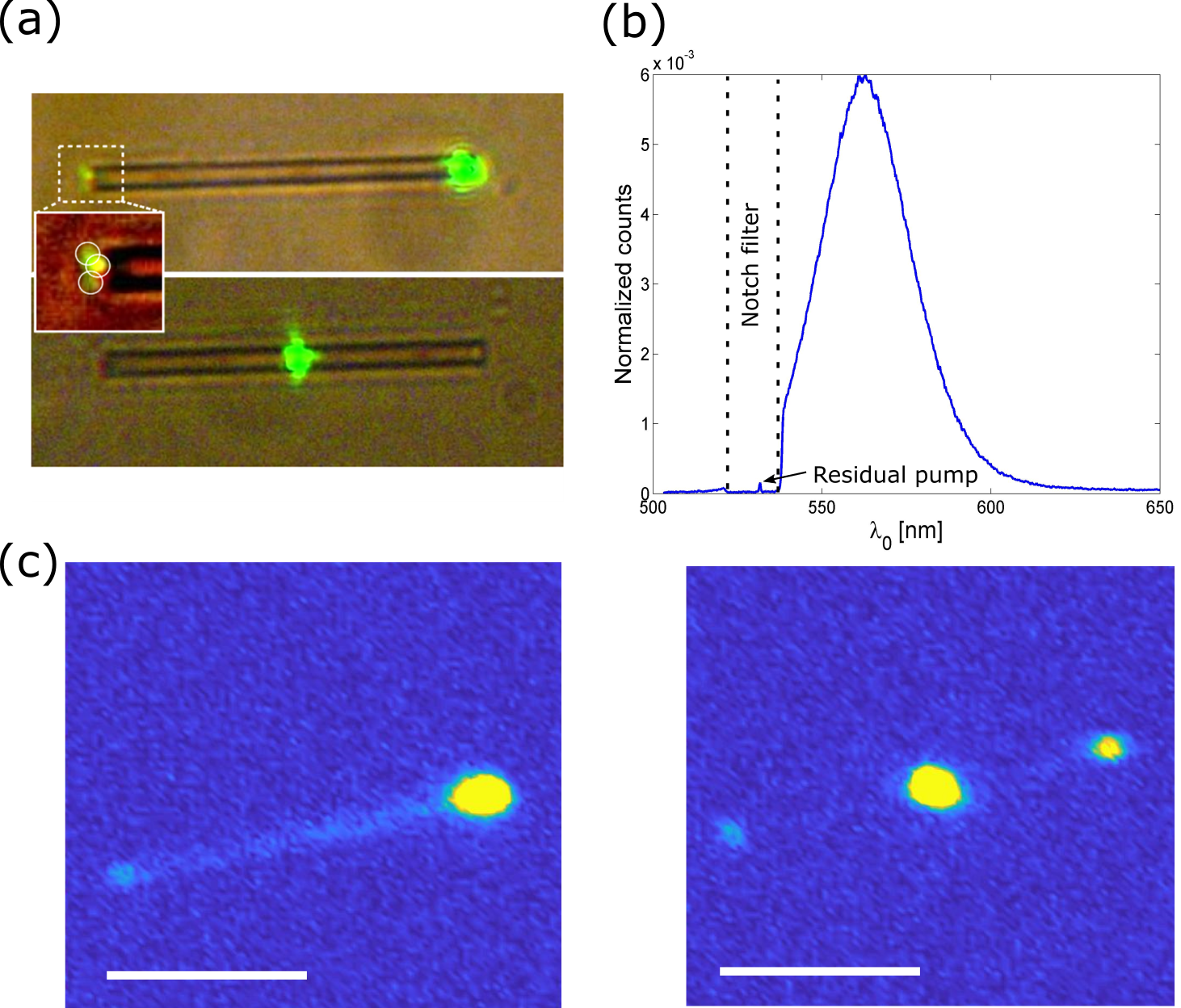}
}
\caption{Characterization of the system. (a) A white light microscope image of the waveguide when green laser light is focused onto the end (upper image) or the middle (lower image) of the waveguides. (b) Photoluminescence spectrum of the CdSe colloidal quantum dots. (c) Confocal map of the system under green light excitation where the excitation is performed onto the end (left image) or the middle (right image) of the waveguide. The scale bars are $8\,\mu m$ both.
\label{Fig3}
}
\end{figure}

The waveguides were fabricated by depositing a layer of SiN on a fused silica substrate using low power chemical vapor deposition, followed by electron beam lithography and dry etching. Scanning electron microscope images of the fabricated slanted waveguide structures are shown in Fig.~\ref{Fig1}(d). The cross sectional dimensions are similar to the one simulated in Fig.~\ref{Fig1}(c) and the length is $15\,\mu m$. We interrogate the system using the confocal microscope schematically shown in Fig.~\ref{Fig2}. For the excitation of the quantum dots we used a continuous wave (cw) or a pulsed laser operating at the wavelength of $532\,nm$ and the fluorescence emission from the sample is collected by two optical channels and detected with avalanche photo diodes. One of the channels is aligned with the excitation laser while the other one can be scanned using galvanometric mirrors, thereby allowing for fluorescence imaging with a fixed excitation spot. We used an oil immersion objective with a numerical aperture of 1.4 to focus and collect the light. For experiments on measuring the quantum dot emission, we inserted a $532\,nm$ notch filter with a $15\,nm$ bandwidth to block the excitation laser.  
We first investigate the propagation of gap modes excited by cw green laser light. As illustrated in Fig.~\ref{Fig3}(a), we launch the green laser beam into the gap mode by focusing the light onto the end facet of the waveguide system. Due to the structural asymmetry at the end facet, photons are scattered from the laser beam into the waveguide modes and propagate along the system to its distal end at which it scatters the photons out. A zoom of the emitted light indicates three emission spots associated with the $TM_p$ mode of the waveguide system. This experiment verifies the existence and propagation of a gap mode. As also illustrated in Fig.~\ref{Fig3}(a), when shining light onto the middle of the waveguide, the mode was not excited and thus the ends of the waveguide remained dark.    

As a next step we wish to excite emitters inside the gap and explore the coupling to the gap mode. We use core CdSe colloidal quantum dots of size $2.6\,nm-3.0\,nm$ dissolved in toluene and at room temperature emitting light in a broad spectrum with a center wavelength around $565\,nm$ (see emission spectrum in Fig.~\ref{Fig3}(b)). The dot solution was ultra-sonicated for 2 minutes before being spin coated onto an ultra clean waveguide sample. We used $20\,\mu L$ of the quantum dot solution while spinning with a speed of $1000\,rpm$ which was increased to $8000\,rpm$ after $10\,s$. The final sample contained an ensemble of quantum dots continuously distributed along the gap, as confirmed by a confocal scan of the system.



\begin{figure}[h]
\centering 
{
\includegraphics[width=0.95\linewidth]
{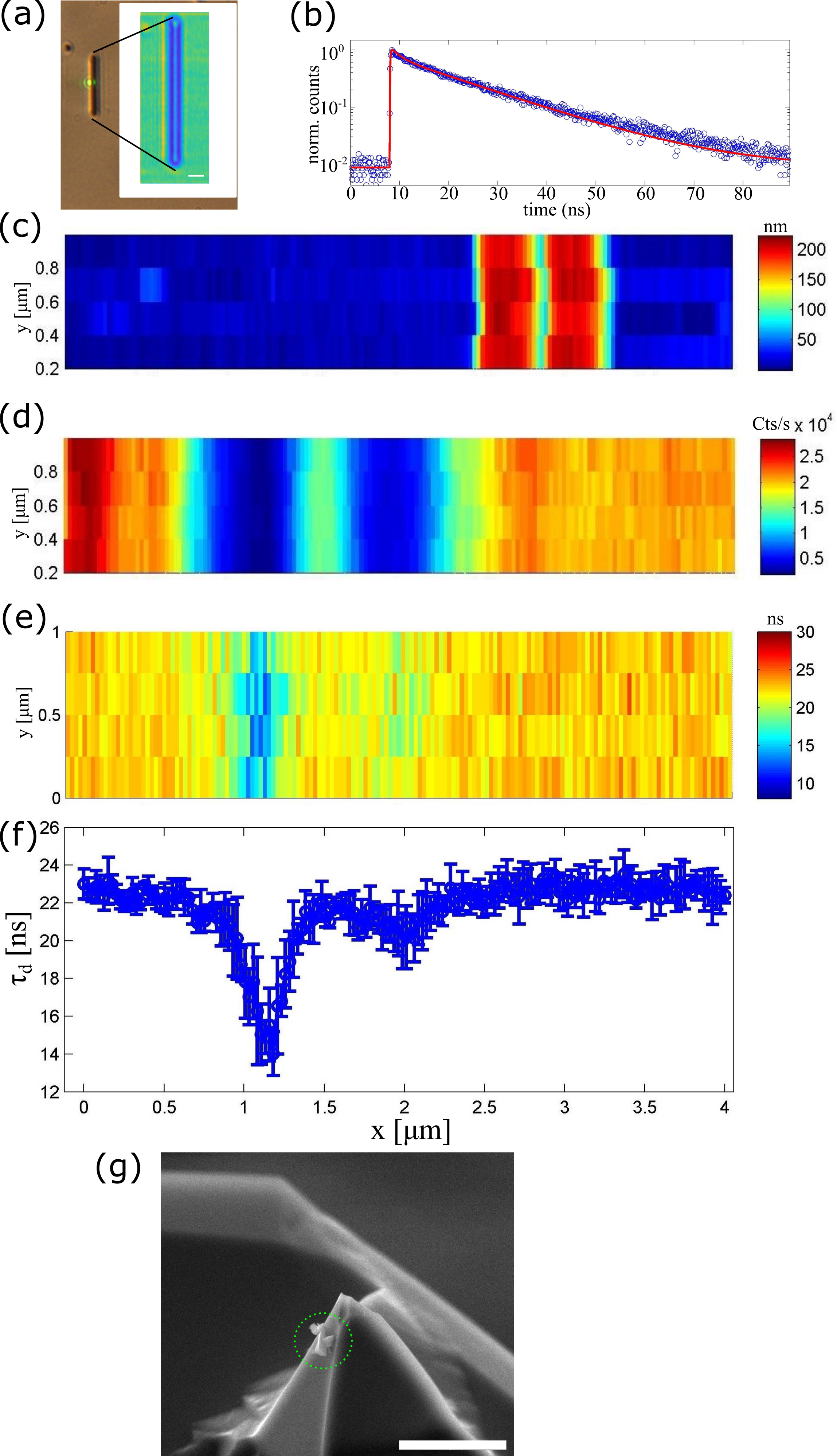}
}
\caption{Characterization of the hybrid quantum dot-waveguide system using the scanning probe technique. (a) An optical image of the entire waveguide structure. Inset: A fluorescence image of the same system. (b) Fluorescence decay curve, where blue circles are experimental data while the red curve is a fit with a multi-explonential function with life time $\tau_d$. (c) AFM height measurement of the waveguide. (d) A fluorescence image of the quantum dot while it is scanned across the waveguide. (e) Lifetime measurements of the quantum dots as they are scanned across the waveguide. (f) Lifetime measurement averaged along the y-direction in (e). (g) A scanning electron microscope image of the cantilever tip with quantum dots on the side of the tip indicated by the green circle. The scale bar is $5\,\mu m$.
\label{Fig4}
}
\end{figure}

The confocal scans presented in Fig.~\ref{Fig3}(c) illustrate our results on coupling emitters to the gap mode. We excite the emitters either at the end or in the middle of the waveguide structure using the cw green excitation beam while scanning an area of $20\,\mu m\times 20\,\mu m$ from which the fluorescence light is measured (the green light was filtered out in this experiment using a notch filter). It is clear that some of the fluorescence light is captured by the waveguide gap modes, propagates along the gap and is re-emitted at the ends. It is also clear that when we excite the emitters at the end of the waveguide, a part of the green light is also launched into the gap mode and thus emitters are excited along the gap. This effect is not significant when exciting the emitters at the center of the waveguide as in this case the excitation beam is not coupled into the waveguide. It means that it is possible to convert the excitation beam into a propagating waveguide mode before it excites the emitters at a location that is remote to the excitation location. 
To investigate whether the dynamics of the emitters is changed due to the coupling to the waveguide mode, we conducted some lifetime measurements of the emitters. However, since the density of emitters in the gap was large, we could not conclude from the acquired measurements whether a change in lifetime is due to coupling to the actual gap mode or simply due to the high density of quantum dots \cite{Kagan1996,Kawata2011,Hartmann2012}. To avoid these density-dependent effects we performed measurements with a few emitters using a scanning-probe fluorescence lifetime imaging microscope~\cite{Michaelis2000,Schell2014}, by which an image of the emitter's lifetimes across the waveguide system could be constructed with nanometer scaled spatial resolution. We include the functionality in our setup by coaxially aligning the silicon cantilever tip of an atomic force microscope (AFM, NT-MDT SMENA) to the focused excitation laser in our confocal microscope, as schematically illustrated in Fig.~\ref{Fig2}. In our realization, we scanned the waveguide sample in the transverse plane while operating the AFM in tapping mode and collected fluorescence photons through the objective towards the apparatus measuring the fluorescence counts and the lifetime. With this technique we were able to make a nanoscaled map of the lifetimes associated with a few emitter attached to the AFM tip in varying environments in which the local optical density of states changes.
Emitters were deposited onto the cantilever tip by dragging it in contact mode through a region with a high concentration of CdSe quantum dots on a reference sample. After removing the AFM cantilever from the region of quantum dots, we observed emitter fluorescence when exciting the tip alone, thus confirming that a few quantum dots have been picked up. Using this tip and changing back to the waveguide sample cleaned with de-ionized water, we produced a real image and a fluorescence image of the waveguides as shown in Fig.~\ref{Fig4}(a). We note that the count rate is much lower atop the waveguides than at the substrate base. This observation we attribute to the fact that in the close vicinity the waveguides modify the quntum dot emission and lower the coupling to the objective. 

We next investigate the potential coupling of the quantum dots to the waveguides by exploring the lifetimes of the quantum dots across the waveguide structure. Towards this end, we carry out a $1\,\mu m\times 4\,\mu m$ scan of the quantum dots across the waveguide system and simultaneously record a topographic, a fluorescence and a lifetime image. The results are illustrated in Fig.\ref{Fig4}(c-f). Note that these measurements are performed with the AFM operating in tapping mode which means that the cantilever oscillates with an amplitude of approximately $70\,nm$ (with a frequency of $178\,kHz$) in a direction orthogonal to the scanning plane. The measured results are therefore averaged over these oscillations. We also note from the measured results in Fig.~\ref{Fig4}(c-d) that the fluorescence image was displaced from the topographic image by approximately $1\,\mu m$. This is caused by the fact that the quantum dots were not attached exactly at the apex of the cantilever tip but at the side of the tip (which was confirmed by a further investigation with scanning electron microscope, see Fig.~\ref{Fig4}(g)). 

To estimate the lifetimes, we simulate the decay mechanism by employing a three level model which accounts for phonon decays followed by photonic decay. The decay pattern could be modelled as \cite{Santori2010}
\begin{equation}
    I_{tot}(t)=I_1(t)+I_2(t)+I_{offset}, \label{eq}
\end{equation}
where 
\begin{equation}
    I_i(t)=A_i\frac{k_{ph,i}k_i}{-k_i+k_{ph,i}}(-e^{-(t-\tau_0)/\tau_{ph,i}}+e^{-(t-\tau_0)/\tau_i}).
\end{equation}
Here, $A_i$ is the amplitude, $\tau_i=1/k_i$ and $\tau_{ph,i}=1/k_{ph,i}$ are the lifetimes associated with the photonic and phononic decays, respectively. The lifetime for the system was then defined as $I_{tot}(\tau_d)=e^{-1}I_{tot}(t=\tau_0)$. The offset, $I_{offset}$, is found by averaging the counts before the arrival of the excitation pulse at $\tau_0$. Fig.~\ref{Fig4}(b) shows an example of a lifetime measurement where we have fitted the multi-exponential function in Eq.~(\ref{eq}) to the experimental data~\cite{Salman2009,Knowles2011}.
In Fig.~\ref{Fig4}(e) and (f) we plot the measured lifetimes, $\tau_d$, across the waveguide structure. It is clear that the lifetime and thus the quantum dot dynamics is modified due to the presence of the waveguides. The lifetime is decreased by up to a factor of approximately 1.5, thus indicating a significant coupling to the waveguide. However, it is also clear that the strongest coupling does not occur in the gap between the waveguides as expected. The missing coupling in the gap is due to the fact that the cantilever blunted by the multiple scans performed throughout the experiment (see Fig.~\ref{Fig4}(g)) and that the quantum dots are positioned at the side of the tip and hence the quantum dots may not go into the gap during the scan. On the other hand, the quantum dots will get very close to the outer wall of one of the waveguides as the cantilever is scanned across. This explains the asymmetry of the lifetime scan in Fig.~\ref{Fig4}(e-f). Therefore, although we have not experimentally proven a strong coupling between the emitter and the gap mode, the emitter is clearly coupling significantly to the waveguiding system. This indicates that a single quantum dot embedded into the gap will be efficiently coupled to the gap mode. 
To observe a change in the lifetime of the quantum dots coupled to the gap mode, it is critical to be able to attach the quantum dots exactly at the apex of the cantilever tip. This turned out to be a huge challenge with the current system since we have no means of observing in real time the attaching process. An effective solution that will facilitate the process and measurement is to combine the AFM and the confocal microscope with a scanning electron microscope. This would allow real-time imaging of the attaching process and at the same time monitoring the fluorescence from the quantum dots. 
In conclusion, we have explored the interaction of quantum dots with modes of a photonic waveguide system. In particular, we have demonstrated the excitation of propagating waveguide modes by quantum dots, and we have witnessed an efficient coupling between the two systems. As an outlook, we suggest to improve the controllability in assembling the system, to reduce the size of the gap for increasing the coupling rate and to test the system with single stable emitters such as single quantum dots or color centers in diamond. Once developed, the platform can potentially serve as a building block for scalable nanoscale photonics~\cite{Harris2016}, quantum dot based bio-analysis~\cite{Altintas2017} and scalable quantum information processing~\cite{Slussarenko2019}.
\begin{acknowledgments}
We greatly acknowledge funding from the Danish Research Council through a Sapere Aude grant
(DIMS, Grant No. 4181-00505B).  
\end{acknowledgments}
\bibliography{References}

\begin{thebibliography}{30}
\expandafter\ifx\csname natexlab\endcsname\relax\def\natexlab#1{#1}\fi
\expandafter\ifx\csname bibnamefont\endcsname\relax
  \def\bibnamefont#1{#1}\fi
\expandafter\ifx\csname bibfnamefont\endcsname\relax
  \def\bibfnamefont#1{#1}\fi
\expandafter\ifx\csname citenamefont\endcsname\relax
  \def\citenamefont#1{#1}\fi
\expandafter\ifx\csname url\endcsname\relax
  \def\url#1{\texttt{#1}}\fi
\expandafter\ifx\csname urlprefix\endcsname\relax\def\urlprefix{URL }\fi
\providecommand{\bibinfo}[2]{#2}
\providecommand{\eprint}[2][]{\url{#2}}

\bibitem[{\citenamefont{Kim et~al.}(2013)\citenamefont{Kim, Bose, Shen,
  Solomon, and Waks}}]{kim2013}
\bibinfo{author}{\bibfnamefont{H.}~\bibnamefont{Kim}},
  \bibinfo{author}{\bibfnamefont{R.}~\bibnamefont{Bose}},
  \bibinfo{author}{\bibfnamefont{T.~C.} \bibnamefont{Shen}},
  \bibinfo{author}{\bibfnamefont{G.~S.} \bibnamefont{Solomon}},
  \bibnamefont{and} \bibinfo{author}{\bibfnamefont{E.}~\bibnamefont{Waks}},
  \bibinfo{journal}{Nature Photonics} \textbf{\bibinfo{volume}{7}},
  \bibinfo{pages}{373} (\bibinfo{year}{2013}), ISSN \bibinfo{issn}{17494885}.

\bibitem[{\citenamefont{Hausmann et~al.}(2012)\citenamefont{Hausmann, Shields,
  Quan, Maletinsky, McCutcheon, Choy, Babinec, Kubanek, Yacoby, Lukin
  et~al.}}]{Hausmann2012}
\bibinfo{author}{\bibfnamefont{B.~J.} \bibnamefont{Hausmann}},
  \bibinfo{author}{\bibfnamefont{B.}~\bibnamefont{Shields}},
  \bibinfo{author}{\bibfnamefont{Q.}~\bibnamefont{Quan}},
  \bibinfo{author}{\bibfnamefont{P.}~\bibnamefont{Maletinsky}},
  \bibinfo{author}{\bibfnamefont{M.}~\bibnamefont{McCutcheon}},
  \bibinfo{author}{\bibfnamefont{J.~T.} \bibnamefont{Choy}},
  \bibinfo{author}{\bibfnamefont{T.~M.} \bibnamefont{Babinec}},
  \bibinfo{author}{\bibfnamefont{A.}~\bibnamefont{Kubanek}},
  \bibinfo{author}{\bibfnamefont{A.}~\bibnamefont{Yacoby}},
  \bibinfo{author}{\bibfnamefont{M.~D.} \bibnamefont{Lukin}},
  \bibnamefont{et~al.}, \bibinfo{journal}{Nano Letters}
  \textbf{\bibinfo{volume}{12}}, \bibinfo{pages}{1578} (\bibinfo{year}{2012}),
  ISSN \bibinfo{issn}{15306984}, \eprint{1111.5330}.

\bibitem[{\citenamefont{Evans et~al.}(2018)\citenamefont{Evans, Bhaskar,
  Sukachev, Nguyen, Sipahigil, Burek, Machielse, Zhang, Zibrov, Bielejec
  et~al.}}]{Evans2018}
\bibinfo{author}{\bibfnamefont{R.~E.} \bibnamefont{Evans}},
  \bibinfo{author}{\bibfnamefont{M.~K.} \bibnamefont{Bhaskar}},
  \bibinfo{author}{\bibfnamefont{D.~D.} \bibnamefont{Sukachev}},
  \bibinfo{author}{\bibfnamefont{C.~T.} \bibnamefont{Nguyen}},
  \bibinfo{author}{\bibfnamefont{A.}~\bibnamefont{Sipahigil}},
  \bibinfo{author}{\bibfnamefont{M.~J.} \bibnamefont{Burek}},
  \bibinfo{author}{\bibfnamefont{B.}~\bibnamefont{Machielse}},
  \bibinfo{author}{\bibfnamefont{G.~H.} \bibnamefont{Zhang}},
  \bibinfo{author}{\bibfnamefont{A.~S.} \bibnamefont{Zibrov}},
  \bibinfo{author}{\bibfnamefont{E.}~\bibnamefont{Bielejec}},
  \bibnamefont{et~al.}, \bibinfo{journal}{Science}
  \textbf{\bibinfo{volume}{362}}, \bibinfo{pages}{662} (\bibinfo{year}{2018}),
  ISSN \bibinfo{issn}{10959203}, \eprint{1807.04265}.

\bibitem[{\citenamefont{Benson}(2011)}]{Benson2011}
\bibinfo{author}{\bibfnamefont{O.}~\bibnamefont{Benson}},
  \bibinfo{journal}{Nature} \textbf{\bibinfo{volume}{480}},
  \bibinfo{pages}{193} (\bibinfo{year}{2011}), ISSN \bibinfo{issn}{00280836},
  \urlprefix\url{http://dx.doi.org/10.1038/nature10610}.

\bibitem[{\citenamefont{Akimov et~al.}(2007)\citenamefont{Akimov, Mukherjee,
  Yu, Chang, Zibrov, Hemmer, Park, and Lukin}}]{Akimov2007}
\bibinfo{author}{\bibfnamefont{A.~V.} \bibnamefont{Akimov}},
  \bibinfo{author}{\bibfnamefont{A.}~\bibnamefont{Mukherjee}},
  \bibinfo{author}{\bibfnamefont{C.~L.} \bibnamefont{Yu}},
  \bibinfo{author}{\bibfnamefont{D.~E.} \bibnamefont{Chang}},
  \bibinfo{author}{\bibfnamefont{A.~S.} \bibnamefont{Zibrov}},
  \bibinfo{author}{\bibfnamefont{P.~R.} \bibnamefont{Hemmer}},
  \bibinfo{author}{\bibfnamefont{H.}~\bibnamefont{Park}}, \bibnamefont{and}
  \bibinfo{author}{\bibfnamefont{M.~D.} \bibnamefont{Lukin}},
  \bibinfo{journal}{Nature} \textbf{\bibinfo{volume}{450}},
  \bibinfo{pages}{402} (\bibinfo{year}{2007}), ISSN \bibinfo{issn}{14764687}.

\bibitem[{\citenamefont{Huck et~al.}((2011))\citenamefont{Huck, Kumar, Shakoor,
  and Andersen}}]{Huck2011}
\bibinfo{author}{\bibfnamefont{A.}~\bibnamefont{Huck}},
  \bibinfo{author}{\bibfnamefont{S.}~\bibnamefont{Kumar}},
  \bibinfo{author}{\bibfnamefont{A.}~\bibnamefont{Shakoor}}, \bibnamefont{and}
  \bibinfo{author}{\bibfnamefont{U.~L.} \bibnamefont{Andersen}},
  \bibinfo{journal}{Phys. Rev. Lett.} \textbf{\bibinfo{volume}{106}},
  \bibinfo{pages}{096801} (\bibinfo{year}{(2011)}).

\bibitem[{\citenamefont{Kumar et~al.}((2013))\citenamefont{Kumar, Huck, and
  Andersen}}]{Kumar2013}
\bibinfo{author}{\bibfnamefont{S.}~\bibnamefont{Kumar}},
  \bibinfo{author}{\bibfnamefont{A.}~\bibnamefont{Huck}}, \bibnamefont{and}
  \bibinfo{author}{\bibfnamefont{U.~L.} \bibnamefont{Andersen}},
  \bibinfo{journal}{Nano Lett.} \textbf{\bibinfo{volume}{13}},
  \bibinfo{pages}{1221} (\bibinfo{year}{(2013)}).

\bibitem[{\citenamefont{Akselrod et~al.}((2014))\citenamefont{Akselrod,
  Argyropoulos, Hoang, ciraci, Fang, Huang, Smith, and
  Mikkelsen}}]{Akselrod2014}
\bibinfo{author}{\bibfnamefont{G.~M.} \bibnamefont{Akselrod}},
  \bibinfo{author}{\bibfnamefont{C.}~\bibnamefont{Argyropoulos}},
  \bibinfo{author}{\bibfnamefont{T.~B.} \bibnamefont{Hoang}},
  \bibinfo{author}{\bibfnamefont{C.}~\bibnamefont{ciraci}},
  \bibinfo{author}{\bibfnamefont{C.}~\bibnamefont{Fang}},
  \bibinfo{author}{\bibfnamefont{J.}~\bibnamefont{Huang}},
  \bibinfo{author}{\bibfnamefont{D.~R.} \bibnamefont{Smith}}, \bibnamefont{and}
  \bibinfo{author}{\bibfnamefont{M.~H.} \bibnamefont{Mikkelsen}},
  \bibinfo{journal}{Nat. Phot.} \textbf{\bibinfo{volume}{8}},
  \bibinfo{pages}{835} (\bibinfo{year}{(2014)}).

\bibitem[{\citenamefont{Kumar et~al.}(2014)\citenamefont{Kumar, Kristiansen,
  Huck, and Andersen}}]{Kumar2014}
\bibinfo{author}{\bibfnamefont{S.}~\bibnamefont{Kumar}},
  \bibinfo{author}{\bibfnamefont{N.~I.} \bibnamefont{Kristiansen}},
  \bibinfo{author}{\bibfnamefont{A.}~\bibnamefont{Huck}}, \bibnamefont{and}
  \bibinfo{author}{\bibfnamefont{U.~L.} \bibnamefont{Andersen}},
  \bibinfo{journal}{Nano Letters} \textbf{\bibinfo{volume}{14}},
  \bibinfo{pages}{663} (\bibinfo{year}{2014}), \bibinfo{note}{pMID: 24471714},
  \eprint{https://doi.org/10.1021/nl403907w},
  \urlprefix\url{https://doi.org/10.1021/nl403907w}.

\bibitem[{\citenamefont{Kumar and Bozhevolnyi}(2019)}]{Kumar2019}
\bibinfo{author}{\bibfnamefont{S.}~\bibnamefont{Kumar}} \bibnamefont{and}
  \bibinfo{author}{\bibfnamefont{S.}~\bibnamefont{Bozhevolnyi}},
  \bibinfo{journal}{A C S Photonics} \textbf{\bibinfo{volume}{6}},
  \bibinfo{pages}{1587} (\bibinfo{year}{2019}), ISSN \bibinfo{issn}{2330-4022}.

\bibitem[{\citenamefont{T{\"{u}}rschmann
  et~al.}(2017)\citenamefont{T{\"{u}}rschmann, Rotenberg, Renger, Harder,
  Lohse, Utikal, G{\"{o}}tzinger, and Sandoghdar}}]{Turschmann2017}
\bibinfo{author}{\bibfnamefont{P.}~\bibnamefont{T{\"{u}}rschmann}},
  \bibinfo{author}{\bibfnamefont{N.}~\bibnamefont{Rotenberg}},
  \bibinfo{author}{\bibfnamefont{J.}~\bibnamefont{Renger}},
  \bibinfo{author}{\bibfnamefont{I.}~\bibnamefont{Harder}},
  \bibinfo{author}{\bibfnamefont{O.}~\bibnamefont{Lohse}},
  \bibinfo{author}{\bibfnamefont{T.}~\bibnamefont{Utikal}},
  \bibinfo{author}{\bibfnamefont{S.}~\bibnamefont{G{\"{o}}tzinger}},
  \bibnamefont{and}
  \bibinfo{author}{\bibfnamefont{V.}~\bibnamefont{Sandoghdar}},
  \bibinfo{journal}{Nano Letters} \textbf{\bibinfo{volume}{17}},
  \bibinfo{pages}{4941} (\bibinfo{year}{2017}), ISSN \bibinfo{issn}{15306992}.

\bibitem[{\citenamefont{Khasminskaya et~al.}(2016)\citenamefont{Khasminskaya,
  Pyatkov, S{\l}owik, Ferrari, Kahl, Kovalyuk, Rath, Vetter, Hennrich, Kappes
  et~al.}}]{Khasminskaya2016}
\bibinfo{author}{\bibfnamefont{S.}~\bibnamefont{Khasminskaya}},
  \bibinfo{author}{\bibfnamefont{F.}~\bibnamefont{Pyatkov}},
  \bibinfo{author}{\bibfnamefont{K.}~\bibnamefont{S{\l}owik}},
  \bibinfo{author}{\bibfnamefont{S.}~\bibnamefont{Ferrari}},
  \bibinfo{author}{\bibfnamefont{O.}~\bibnamefont{Kahl}},
  \bibinfo{author}{\bibfnamefont{V.}~\bibnamefont{Kovalyuk}},
  \bibinfo{author}{\bibfnamefont{P.}~\bibnamefont{Rath}},
  \bibinfo{author}{\bibfnamefont{A.}~\bibnamefont{Vetter}},
  \bibinfo{author}{\bibfnamefont{F.}~\bibnamefont{Hennrich}},
  \bibinfo{author}{\bibfnamefont{M.~M.} \bibnamefont{Kappes}},
  \bibnamefont{et~al.}, \bibinfo{journal}{Nature Photonics}
  \textbf{\bibinfo{volume}{10}}, \bibinfo{pages}{727} (\bibinfo{year}{2016}),
  ISSN \bibinfo{issn}{17494893}.

\bibitem[{\citenamefont{Davanco et~al.}(2017)\citenamefont{Davanco, Liu,
  Sapienza, Zhang, {De Miranda Cardoso}, Verma, Mirin, Nam, Liu, and
  Srinivasan}}]{Davanco2017}
\bibinfo{author}{\bibfnamefont{M.}~\bibnamefont{Davanco}},
  \bibinfo{author}{\bibfnamefont{J.}~\bibnamefont{Liu}},
  \bibinfo{author}{\bibfnamefont{L.}~\bibnamefont{Sapienza}},
  \bibinfo{author}{\bibfnamefont{C.~Z.} \bibnamefont{Zhang}},
  \bibinfo{author}{\bibfnamefont{J.~V.} \bibnamefont{{De Miranda Cardoso}}},
  \bibinfo{author}{\bibfnamefont{V.}~\bibnamefont{Verma}},
  \bibinfo{author}{\bibfnamefont{R.}~\bibnamefont{Mirin}},
  \bibinfo{author}{\bibfnamefont{S.~W.} \bibnamefont{Nam}},
  \bibinfo{author}{\bibfnamefont{L.}~\bibnamefont{Liu}}, \bibnamefont{and}
  \bibinfo{author}{\bibfnamefont{K.}~\bibnamefont{Srinivasan}},
  \bibinfo{journal}{Nature Communications} \textbf{\bibinfo{volume}{8}},
  \bibinfo{pages}{1} (\bibinfo{year}{2017}), ISSN \bibinfo{issn}{20411723},
  \eprint{1611.07654},
  \urlprefix\url{http://dx.doi.org/10.1038/s41467-017-00987-6}.

\bibitem[{\citenamefont{Elshaari et~al.}(2017)\citenamefont{Elshaari, Zadeh,
  Fognini, Reimer, Dalacu, Poole, Zwiller, and J{\"{o}}ns}}]{Elshaari2017}
\bibinfo{author}{\bibfnamefont{A.~W.} \bibnamefont{Elshaari}},
  \bibinfo{author}{\bibfnamefont{I.~E.} \bibnamefont{Zadeh}},
  \bibinfo{author}{\bibfnamefont{A.}~\bibnamefont{Fognini}},
  \bibinfo{author}{\bibfnamefont{M.~E.} \bibnamefont{Reimer}},
  \bibinfo{author}{\bibfnamefont{D.}~\bibnamefont{Dalacu}},
  \bibinfo{author}{\bibfnamefont{P.~J.} \bibnamefont{Poole}},
  \bibinfo{author}{\bibfnamefont{V.}~\bibnamefont{Zwiller}}, \bibnamefont{and}
  \bibinfo{author}{\bibfnamefont{K.~D.} \bibnamefont{J{\"{o}}ns}},
  \bibinfo{journal}{Nature Communications} \textbf{\bibinfo{volume}{8}},
  \bibinfo{pages}{1} (\bibinfo{year}{2017}), ISSN \bibinfo{issn}{20411723},
  \urlprefix\url{http://dx.doi.org/10.1038/s41467-017-00486-8}.

\bibitem[{\citenamefont{Kim et~al.}(2017)\citenamefont{Kim, Aghaeimeibodi,
  Richardson, Leavitt, Englund, and Waks}}]{kim2017}
\bibinfo{author}{\bibfnamefont{J.~H.} \bibnamefont{Kim}},
  \bibinfo{author}{\bibfnamefont{S.}~\bibnamefont{Aghaeimeibodi}},
  \bibinfo{author}{\bibfnamefont{C.~J.} \bibnamefont{Richardson}},
  \bibinfo{author}{\bibfnamefont{R.~P.} \bibnamefont{Leavitt}},
  \bibinfo{author}{\bibfnamefont{D.}~\bibnamefont{Englund}}, \bibnamefont{and}
  \bibinfo{author}{\bibfnamefont{E.}~\bibnamefont{Waks}},
  \bibinfo{journal}{Nano Letters} \textbf{\bibinfo{volume}{17}},
  \bibinfo{pages}{7394} (\bibinfo{year}{2017}), ISSN \bibinfo{issn}{15306992}.

\bibitem[{\citenamefont{Osada et~al.}(2019)\citenamefont{Osada, Ota, Katsumi,
  Kakuda, Iwamoto, and Arakawa}}]{Osada2019}
\bibinfo{author}{\bibfnamefont{A.}~\bibnamefont{Osada}},
  \bibinfo{author}{\bibfnamefont{Y.}~\bibnamefont{Ota}},
  \bibinfo{author}{\bibfnamefont{R.}~\bibnamefont{Katsumi}},
  \bibinfo{author}{\bibfnamefont{M.}~\bibnamefont{Kakuda}},
  \bibinfo{author}{\bibfnamefont{S.}~\bibnamefont{Iwamoto}}, \bibnamefont{and}
  \bibinfo{author}{\bibfnamefont{Y.}~\bibnamefont{Arakawa}},
  \bibinfo{journal}{Physical Review Applied} \textbf{\bibinfo{volume}{11}},
  \bibinfo{pages}{1} (\bibinfo{year}{2019}), ISSN \bibinfo{issn}{23317019},
  \eprint{1809.11025},
  \urlprefix\url{https://doi.org/10.1103/PhysRevApplied.11.024071}.

\bibitem[{\citenamefont{Katsumi et~al.}(2019)\citenamefont{Katsumi, Ota, Osada,
  Yamaguchi, Tajiri, Kakuda, Iwamoto, Akiyama, and Arakawa}}]{Katsumi2019}
\bibinfo{author}{\bibfnamefont{R.}~\bibnamefont{Katsumi}},
  \bibinfo{author}{\bibfnamefont{Y.}~\bibnamefont{Ota}},
  \bibinfo{author}{\bibfnamefont{A.}~\bibnamefont{Osada}},
  \bibinfo{author}{\bibfnamefont{T.}~\bibnamefont{Yamaguchi}},
  \bibinfo{author}{\bibfnamefont{T.}~\bibnamefont{Tajiri}},
  \bibinfo{author}{\bibfnamefont{M.}~\bibnamefont{Kakuda}},
  \bibinfo{author}{\bibfnamefont{S.}~\bibnamefont{Iwamoto}},
  \bibinfo{author}{\bibfnamefont{H.}~\bibnamefont{Akiyama}}, \bibnamefont{and}
  \bibinfo{author}{\bibfnamefont{Y.}~\bibnamefont{Arakawa}},
  \bibinfo{journal}{APL Photonics} \textbf{\bibinfo{volume}{4}}
  (\bibinfo{year}{2019}), ISSN \bibinfo{issn}{23780967}, \eprint{1812.11666},
  \urlprefix\url{http://dx.doi.org/10.1063/1.5087263}.

\bibitem[{\citenamefont{Kim et~al.}(2019)\citenamefont{Kim, Duong, Nguyen, Lu,
  Kianinia, Mendelson, Solntsev, Bradac, Englund, and Aharonovich}}]{Kim2019}
\bibinfo{author}{\bibfnamefont{S.}~\bibnamefont{Kim}},
  \bibinfo{author}{\bibfnamefont{N.~M.~H.} \bibnamefont{Duong}},
  \bibinfo{author}{\bibfnamefont{M.}~\bibnamefont{Nguyen}},
  \bibinfo{author}{\bibfnamefont{T.~J.} \bibnamefont{Lu}},
  \bibinfo{author}{\bibfnamefont{M.}~\bibnamefont{Kianinia}},
  \bibinfo{author}{\bibfnamefont{N.}~\bibnamefont{Mendelson}},
  \bibinfo{author}{\bibfnamefont{A.}~\bibnamefont{Solntsev}},
  \bibinfo{author}{\bibfnamefont{C.}~\bibnamefont{Bradac}},
  \bibinfo{author}{\bibfnamefont{D.~R.} \bibnamefont{Englund}},
  \bibnamefont{and}
  \bibinfo{author}{\bibfnamefont{I.}~\bibnamefont{Aharonovich}},
  \bibinfo{journal}{Advanced Optical Materials} \textbf{\bibinfo{volume}{7}},
  \bibinfo{pages}{1} (\bibinfo{year}{2019}), ISSN \bibinfo{issn}{21951071},
  \eprint{1907.04546}.

\bibitem[{\citenamefont{Robinson et~al.}((2005))\citenamefont{Robinson,
  Manolatou, Chen, and Lipson}}]{Robinson2005}
\bibinfo{author}{\bibfnamefont{J.~T.} \bibnamefont{Robinson}},
  \bibinfo{author}{\bibfnamefont{C.}~\bibnamefont{Manolatou}},
  \bibinfo{author}{\bibfnamefont{L.}~\bibnamefont{Chen}}, \bibnamefont{and}
  \bibinfo{author}{\bibfnamefont{M.}~\bibnamefont{Lipson}},
  \bibinfo{journal}{Phys. Rev. Lett.} \textbf{\bibinfo{volume}{95}},
  \bibinfo{pages}{143901} (\bibinfo{year}{(2005)}).

\bibitem[{\citenamefont{Kagan et~al.}((1996))\citenamefont{Kagan, Murray, and
  Bawendi}}]{Kagan1996}
\bibinfo{author}{\bibfnamefont{C.~R.} \bibnamefont{Kagan}},
  \bibinfo{author}{\bibfnamefont{C.~B.} \bibnamefont{Murray}},
  \bibnamefont{and} \bibinfo{author}{\bibfnamefont{M.~G.}
  \bibnamefont{Bawendi}}, \bibinfo{journal}{Phys Rev. B}
  \textbf{\bibinfo{volume}{54}}, \bibinfo{pages}{8633}
  (\bibinfo{year}{(1996)}).

\bibitem[{\citenamefont{Kawata et~al.}((2011))\citenamefont{Kawata, Ogawa, and
  Minami}}]{Kawata2011}
\bibinfo{author}{\bibfnamefont{G.}~\bibnamefont{Kawata}},
  \bibinfo{author}{\bibfnamefont{Y.}~\bibnamefont{Ogawa}}, \bibnamefont{and}
  \bibinfo{author}{\bibfnamefont{F.}~\bibnamefont{Minami}},
  \bibinfo{journal}{Jour. of Appl. Phys.} \textbf{\bibinfo{volume}{110}},
  \bibinfo{pages}{064323} (\bibinfo{year}{(2011)}).

\bibitem[{\citenamefont{Hartmann et~al.}((2012))\citenamefont{Hartmann, Kumar,
  Welker, Fiore, Julen-Rabant, Gromova, Bardet, Reiss, Baxter, Chandezon
  et~al.}}]{Hartmann2012}
\bibinfo{author}{\bibfnamefont{L.}~\bibnamefont{Hartmann}},
  \bibinfo{author}{\bibfnamefont{A.}~\bibnamefont{Kumar}},
  \bibinfo{author}{\bibfnamefont{M.}~\bibnamefont{Welker}},
  \bibinfo{author}{\bibfnamefont{A.}~\bibnamefont{Fiore}},
  \bibinfo{author}{\bibfnamefont{C.}~\bibnamefont{Julen-Rabant}},
  \bibinfo{author}{\bibfnamefont{M.}~\bibnamefont{Gromova}},
  \bibinfo{author}{\bibfnamefont{M.}~\bibnamefont{Bardet}},
  \bibinfo{author}{\bibfnamefont{P.}~\bibnamefont{Reiss}},
  \bibinfo{author}{\bibfnamefont{P.~N.~W.} \bibnamefont{Baxter}},
  \bibinfo{author}{\bibfnamefont{F.}~\bibnamefont{Chandezon}},
  \bibnamefont{et~al.}, \bibinfo{journal}{ACS Nano}
  \textbf{\bibinfo{volume}{6}}, \bibinfo{pages}{9033} (\bibinfo{year}{(2012)}).

\bibitem[{\citenamefont{J.~Michaelis and Sandoghdar}((2000))}]{Michaelis2000}
\bibinfo{author}{\bibfnamefont{J.~M.} \bibnamefont{J.~Michaelis},
  \bibfnamefont{C.~Hettich}} \bibnamefont{and}
  \bibinfo{author}{\bibfnamefont{V.}~\bibnamefont{Sandoghdar}},
  \bibinfo{journal}{Nature} \textbf{\bibinfo{volume}{405}},
  \bibinfo{pages}{325} (\bibinfo{year}{(2000)}).

\bibitem[{\citenamefont{Schell et~al.}((2014))\citenamefont{Schell, Engel,
  Werra, Wolff, Busch, and Benson}}]{Schell2014}
\bibinfo{author}{\bibfnamefont{A.~W.} \bibnamefont{Schell}},
  \bibinfo{author}{\bibfnamefont{P.}~\bibnamefont{Engel}},
  \bibinfo{author}{\bibfnamefont{J.~F.~M.} \bibnamefont{Werra}},
  \bibinfo{author}{\bibfnamefont{C.}~\bibnamefont{Wolff}},
  \bibinfo{author}{\bibfnamefont{K.}~\bibnamefont{Busch}}, \bibnamefont{and}
  \bibinfo{author}{\bibfnamefont{O.}~\bibnamefont{Benson}},
  \bibinfo{journal}{Nano Lett.} \textbf{\bibinfo{volume}{14}},
  \bibinfo{pages}{2623} (\bibinfo{year}{(2014)}).

\bibitem[{\citenamefont{C.~Santori and Yamamoto}((2010))}]{Santori2010}
\bibinfo{author}{\bibfnamefont{D.~F.} \bibnamefont{C.~Santori}}
  \bibnamefont{and} \bibinfo{author}{\bibfnamefont{Y.}~\bibnamefont{Yamamoto}},
  \bibinfo{journal}{Wiley-VCH}  (\bibinfo{year}{(2010)}).

\bibitem[{\citenamefont{Salman et~al.}((2009))\citenamefont{Salman,
  Tortschanoff, van~der Zwan, van Mourik, and Chergui}}]{Salman2009}
\bibinfo{author}{\bibfnamefont{A.~A.} \bibnamefont{Salman}},
  \bibinfo{author}{\bibfnamefont{A.}~\bibnamefont{Tortschanoff}},
  \bibinfo{author}{\bibfnamefont{G.}~\bibnamefont{van~der Zwan}},
  \bibinfo{author}{\bibfnamefont{F.}~\bibnamefont{van Mourik}},
  \bibnamefont{and} \bibinfo{author}{\bibfnamefont{M.}~\bibnamefont{Chergui}},
  \bibinfo{journal}{Chem. Phys.} \textbf{\bibinfo{volume}{357}},
  \bibinfo{pages}{96} (\bibinfo{year}{(2009)}).

\bibitem[{\citenamefont{Knowles et~al.}((2011))\citenamefont{Knowles, McArthur,
  and Weiss}}]{Knowles2011}
\bibinfo{author}{\bibfnamefont{K.~E.} \bibnamefont{Knowles}},
  \bibinfo{author}{\bibfnamefont{E.~A.} \bibnamefont{McArthur}},
  \bibnamefont{and} \bibinfo{author}{\bibfnamefont{E.~A.} \bibnamefont{Weiss}},
  \bibinfo{journal}{ACS Nano} \textbf{\bibinfo{volume}{5}},
  \bibinfo{pages}{2026} (\bibinfo{year}{(2011)}).

\bibitem[{\citenamefont{Harris et~al.}(2016)\citenamefont{Harris, Bunandar,
  Pant, Steinbrecher, Mower, Prabhu, Baehr-Jones, Hochberg, and
  Englund}}]{Harris2016}
\bibinfo{author}{\bibfnamefont{N.~C.} \bibnamefont{Harris}},
  \bibinfo{author}{\bibfnamefont{D.}~\bibnamefont{Bunandar}},
  \bibinfo{author}{\bibfnamefont{M.}~\bibnamefont{Pant}},
  \bibinfo{author}{\bibfnamefont{G.~R.} \bibnamefont{Steinbrecher}},
  \bibinfo{author}{\bibfnamefont{J.}~\bibnamefont{Mower}},
  \bibinfo{author}{\bibfnamefont{M.}~\bibnamefont{Prabhu}},
  \bibinfo{author}{\bibfnamefont{T.}~\bibnamefont{Baehr-Jones}},
  \bibinfo{author}{\bibfnamefont{M.}~\bibnamefont{Hochberg}}, \bibnamefont{and}
  \bibinfo{author}{\bibfnamefont{D.}~\bibnamefont{Englund}},
  \textbf{\bibinfo{volume}{5}}, \bibinfo{pages}{456} (\bibinfo{year}{2016}),
  ISSN \bibinfo{issn}{21928614}.

\bibitem[{\citenamefont{Altintas et~al.}(2017)\citenamefont{Altintas, Davis,
  and Scheller}}]{Altintas2017}
\bibinfo{author}{\bibfnamefont{Z.}~\bibnamefont{Altintas}},
  \bibinfo{author}{\bibfnamefont{F.}~\bibnamefont{Davis}}, \bibnamefont{and}
  \bibinfo{author}{\bibfnamefont{F.~W.} \bibnamefont{Scheller}},
  \emph{\bibinfo{title}{Applications of Quantum Dots in Biosensors and
  Diagnostics}} (\bibinfo{publisher}{John Wiley \& Sons, Ltd},
  \bibinfo{year}{2017}), chap.~\bibinfo{chapter}{9}, pp.
  \bibinfo{pages}{183--199}, ISBN \bibinfo{isbn}{9781119065036}.

\bibitem[{\citenamefont{Slussarenko and Pryde}(2019)}]{Slussarenko2019}
\bibinfo{author}{\bibfnamefont{S.}~\bibnamefont{Slussarenko}} \bibnamefont{and}
  \bibinfo{author}{\bibfnamefont{G.~J.} \bibnamefont{Pryde}},
  \bibinfo{journal}{Applied Physics Reviews} \textbf{\bibinfo{volume}{6}},
  \bibinfo{pages}{041303} (\bibinfo{year}{2019}),
  \eprint{https://doi.org/10.1063/1.5115814},
  \urlprefix\url{https://doi.org/10.1063/1.5115814}.

\end{thebibliography}
\end{document}